\documentclass[a4paper,11pt]{article}

\usepackage[utf8]{inputenc}
\usepackage[T1,T2A]{fontenc}
\usepackage{amsmath,amsfonts,amssymb}

\usepackage{graphicx}
\usepackage{wrapfig}
\usepackage{subfig}
\usepackage{hyperref}
\usepackage[dvipsnames]{xcolor}

\begin{document}

\begin{center}
{\LARGE {\bf Invisible Higgs decay from dark matter freeze-in at stronger coupling}}
\end{center}

\vspace{1cm}
\begin{center}
{\bf Oleg Lebedev$^{\,1}$, António P. Morais$^{\,2}$, Vinícius Oliveira$^{\,2}$, Roman Pasechnik$^{\,3}$}
\end{center}

\begin{center}
  \vspace*{0.25cm}
 $^1$ \it{Department of Physics and Helsinki Institute of Physics,\\
  Gustaf H\"allstr\"omin katu 2a, FI-00014 Helsinki, Finland}\\
  \vspace{0.2cm}
  $^2$ \it{Departamento de F\'{i}sica da Universidade de Aveiro and \\ Centre  for  Research  and   Development  in  Mathematics  and  Applications  (CIDMA), \\
  Campus de Santiago, 3810-183 Aveiro, Portugal.}\\
  \vspace{0.2cm}
  $^3$ \it{Department of Physics, Lund University, SE-223 62 Lund, Sweden}\\
\end{center}

\vspace{2.5cm}

\begin{center} {\bf Abstract} \end{center}
\noindent
We study the Higgs boson decay into dark matter (DM) in the framework of freeze-in at stronger coupling.  Even though the Higgs-DM coupling 
is significant, up to order one, DM does not thermalize due to the Boltzmann suppression of its production at low temperatures. We find that this 
mechanism leads to observable Higgs decay into invisible final states with the branching fraction of 10\% and below, while producing the correct 
DM relic abundance. This applies to the DM masses down to the MeV scale, which requires a careful treatment of the hadronic production modes. 
For DM masses below the muon threshold, the Boltzmann suppression is not operative and the freeze-in nature of the production mechanism 
is instead guaranteed by the smallness of the electron Yukawa coupling. As a result, MeV DM with a significant coupling to the Higgs boson remains 
non-thermal as long as the reheating temperature does not exceed ${\cal O}(100)\;$MeV. Our findings indicate that there are good prospects 
for observing light non-thermal DM via invisible Higgs decay at the LHC and FCC.

\vspace{1cm}

\newpage 
\tableofcontents

\section{Introduction}
 
The Higgs boson \cite{Higgs:1964pj,Englert:2014zpa} properties serve as a sensitive probe for new physics. Its couplings to the Standard Model (SM) fields 
can be modified by the presence of extra states and, also, additional interactions and decay channels may appear \cite{Bass:2021acr}. In our work, 
we focus on the Higgs decay into light DM, which effectively corresponds to an ``invisible'' final state.

In the Weakly Interacting Massive Particle (WIMP) models, DM reaches thermal equilibrium with the SM bath and its eventual density is determined 
by the freeze-out mechanism. This requires the DM coupling to the SM states to be significant. In the Higgs portal models \cite{Silveira:1985rk,Patt:2006fw}, 
the coupling to the Higgs field has to be large enough, and if DM is light the consequent invisible Higgs decay would be very efficient. On the other hand, 
the Higgs data from the Large Hadron Collider (LHC) require the branching ratio for such a decay mode to be below 10\% or so. This essentially rules out 
light thermal DM in the Higgs portal framework \cite{Djouadi:2011aa}.

DM can also be non-thermal, while being produced by the SM thermal bath \cite{Dodelson:1993je}. This mechanism is known as ``freeze-in'' \cite{Hall:2009bx}. 
The coupling to the SM sector is feeble such that DM production is slow enough, which also implies that there are no observable signatures of this scenario at colliders. 

The recently proposed option of ``freeze-in at stronger coupling'' \cite{Cosme:2023xpa} evades this conclusion \cite{Koivunen:2024vhr,Arcadi:2024wwg}. In this case, 
the reheating temperature $T_R$ is lower than the DM mass $m_{\rm DM}$, which suppresses its production and thermalization even at larger coupling. As a result, 
the Higgs-DM interaction can be relatively strong leading to observable Higgs decay into DM as well as direct DM detection prospects. This motivates further 
improvements in the Higgs precision measurements.

The low $T_R$ models are also motivated from the top-down perspective. The traditional freeze-in models suffer from the gravitational DM production background
\cite{Lebedev:2022cic,Koutroulis:2023fgp}. All particles are produced during and immediately after inflation via their gravitational couplings, be it classical or quantum 
in nature. While the fields with significant couplings decay or thermalize, feebly interacting DM remains and creates a background for freeze-in calculations.
However, if reheating happens at late times corresponding to a low $T_R$, the post-inflationary DM abundance gets diluted. One may therefore (justifiably) 
assume a negligible initial density in the low-$T$ freeze-in calculations. 

In this work, we study the Higgs decay into scalar and fermion DM in simplest Higgs portal models. Our main assumption is that the SM bath temperature never 
exceeds a few GeV, which is consistent with the cosmological data. One of the challenges of the low-$T$ freeze-in calculations is posed by hadronization below 
the critical QCD temperature $T_c$. Hadron annihilation provides the leading mode for DM production in a wide parameter range. Although this process is 
non-perturbative, it can be accounted for by relating the reaction rate to the Higgs decay rate into hadrons. We delineate parameter space leading to the correct 
DM relic abundance and assess the prospects of observing the Higgs decay into light DM.

\section{Theoretical setup}
\label{Sect:theory}

The goal of this work is to  study ``invisible'' Higgs decay into DM in the framework of freeze-in at stronger coupling \cite{Cosme:2023xpa}. 
We assume the minimal Higgs portal setup with singlet DM of spin 0 or 1/2 (for a review, see \cite{Lebedev:2021xey}). In what follows, 
we consider these options separately and allow for both CP-even and CP-odd couplings of the DM fermion to the Higgs field.
The interactions between the SM Higgs SU(2)$_{\rm W}$-doublet ${\cal H}$ and DM are given by 
 \begin{eqnarray}
{\cal L}_{hs} &=& {\lambda_{hs} \over 2}\; {\cal H}^\dagger {\cal H} SS  \;, 
\label{Higgs-portal1}
\\
  {\cal L}_{h\chi}&=& {1\over \Lambda} {\cal H}^\dagger {\cal H} \, \bar \chi \chi ~~,~~
  {\cal L}_{h\chi}^{\gamma_5}= {1\over \Lambda_5} {\cal H}^\dagger {\cal H} \, \bar \chi \,i\gamma_5\,\chi \;,
\label{Higgs-portal2}
\end{eqnarray}
where $S$ is a real scalar with mass $m_s$ and $\chi$ is a {\it Majorana} fermion with mass $m_\chi$. These states are assumed to be stable 
and thus constitute DM. In the fermion case  \cite{Kim:2006af,Lopez-Honorez:2012tov}, the coupling to the SM is non-renormalizable and 
thus necessitates introduction of the ``new physics'' scales $\Lambda$, $\Lambda_5$.

We are interested in the energy scales below $m_h/2$, hence the physical Higgs boson $h$ of mass $m_h\simeq 125$ GeV 
can be integrated out for practical purposes. Given the SM fermion $f$ (with mass $m_f$) coupling
 \begin{equation}
{\cal L}_{\rm SM}= {m_f\over v}\; h \bar f f  \;,
\end{equation}
with $v\simeq 246$ GeV being the SM Higgs vacuum expectation value, we obtain the effective interactions 
 \begin{eqnarray}
{\cal L}_{{\rm eff} }^{s} &=& {\lambda_{hs} m_f \over 2 m_h^2}\; \bar f f \, SS  \;, 
\label{Higgs-portal1-eff}
\\
  {\cal L}_{\rm eff}^{\chi}&=& {m_f\over \Lambda m_h^2 } \, \bar f f \, \bar \chi \chi ~~,~~
  {\cal L}_{\rm eff}^{\chi ~\gamma_5}= {m_f\over \Lambda_5 m_h^2 } \, \bar f f \, \bar \chi \, i\gamma_5 \, \chi  \;.
\label{Higgs-portal2-eff}
\end{eqnarray}
These operators are responsible for producing DM from the SM thermal bath at low temperatures. The above Lagrangian also leads to the photon 
coupling to DM, yet we find the effect to be small.
 \begin{figure}[t!]
    \centering
    \includegraphics[width=0.7  \textwidth]{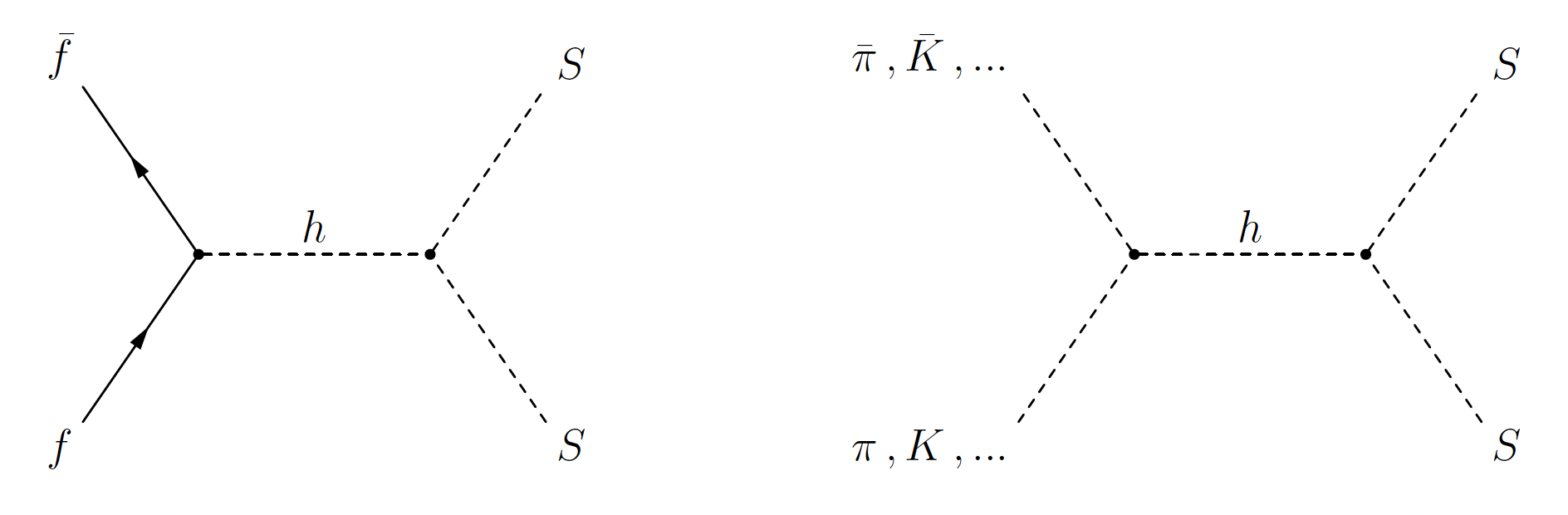}
    \caption{ Main contributions to scalar DM production at low temperatures.}
    \label{diagram}
\end{figure}

In the framework of freeze-in at stronger coupling, DM is produced via annihilation of the SM thermal bath states at low temperatures 
as illustrated in Fig.\,\ref{diagram} in the case of scalar DM. 
If temperature $T$ is never higher than the DM mass, the production rate is Boltzmann-suppressed\footnote{Boltzmann-suppressed production has also 
been considered in the context of super-heavy DM in \cite{Okada:2021uqk} and in the context of phase transitions in \cite{Wong:2023qon}.}, which allows 
for a significant Higgs coupling to DM. The latter can then lead to observable Higgs decay into ``invisible'' final states. The production rate calculation 
is perturbative above the confinement scale, $T \gtrsim 150\;$MeV, while at lower temperatures the processes become complicated due to hadronization. 
The main challenge is to account for initial hadronic states in the plasma. To address this issue, we use a thermodynamic relation between the DM production 
and annihilation rates, while expressing the latter in terms of the Higgs decay width into hadrons,
\begin{equation} 
\Gamma ({\rm hadrons} \rightarrow h \rightarrow {\rm DM}) ~\leftrightarrow ~\Gamma (h \rightarrow {\rm hadrons}) \;.
\end{equation}
Here, $\Gamma (h \rightarrow {\rm hadrons}) $ is considered to be the SM Higgs decay width into hadrons with a {\it variable} Higgs mass corresponding 
to the center-of-mass energy of the DM annihilation process. It has been computed by different groups using various approaches including the chiral 
perturbation theory, dispersion relations, etc. The above relation allows us to include the hadronic modes consistently, albeit with some uncertainty inherent 
in such non-perturbative calculations.

We explore the SM bath temperatures down to $T=4\;$MeV, which is the lowest reheating temperature consistent with observations \cite{Hannestad:2004px}. 
Our main conclusion is that the DM masses above 1 MeV are allowed and, depending on the reheating temperature, this can lead to a significant Higgs invisible 
decay branching ratio, at the level of 10\% and below. As a result, the framework can be probed further at the LHC as well as the Future Circular Collider (FCC).

\section{Scalar dark matter}
\label{Sect:scalarDM}

Consider first the scalar DM case \cite{Silveira:1985rk}. It is produced by the SM thermal bath via the effective coupling $\bar f f \, SS$. The DM number density 
evolution is governed by the Boltzmann equation, which we discuss in detail below.
 
\subsection{Boltzmann equation and the basics}
 
The Boltzmann equation for the DM number density $n$ takes the form
\begin{eqnarray}
\dot n + 3Hn  = 2 \,\Gamma ({\rm SM \rightarrow DM}) - 2\,\Gamma ({\rm DM \rightarrow SM})  \;,
\end{eqnarray}
where $H$ is the Hubble rate, and $\Gamma ({\rm SM \rightarrow DM}),  \Gamma ({\rm DM \rightarrow SM}) $ are the DM production and annihilation rates 
per unit volume, respectively. Here, the factor of two signifies production/annihilation of two DM quanta in each reaction. In the regime of interest, the characteristic 
energy of the process is above the temperature such that the quantum statistics effects can be neglected. In this case, the $a \rightarrow b$ reaction rate per unit 
volume is given by the general expression
\begin{equation}
\Gamma_{a\rightarrow b} = \int \left( \prod_{i\in a} {d^3 {\bf p}_i \over (2 \pi)^3 2E_{i}} f(p_i)\right)~
\left( \prod_{j\in b} {d^3 {\bf p}_j \over (2 \pi)^3 2E_{j}} \right)
\vert {\cal M}_{a\rightarrow b} \vert^2 ~ (2\pi)^4 \delta^4(p_a-p_b) \;,
\label{Gamma}
\end{equation}
where $p_i$ and $p_j$ are the initial and final state momenta, respectively; $f(p)$ is the momentum distribution function, and ${\cal M}_{a\rightarrow b}$ is 
the QFT $a \rightarrow b$ transition amplitude. For convenience, we absorb both the {\it initial and final} state phase space symmetry factors into 
$\vert {\cal M}_{a\rightarrow b} \vert^2$.

\begin{figure}[t!]
    \centering
    \includegraphics[width=0.5 \textwidth]{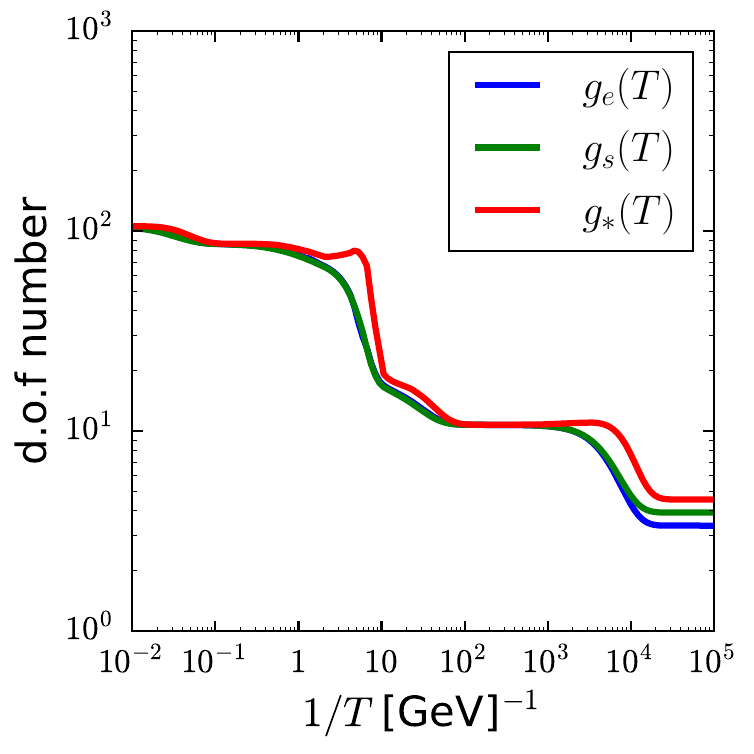}
    \caption{SM d.o.f. number as a function of temperature \cite{Gondolo:1990dk}.}
    \label{dof}
\end{figure}

In what follows, we will primarily be interested in the $pure$ {\it freeze-in} regime, where the annihilation term is negligible. Consider the production 
mode $\bar f f \rightarrow SS$. The DM abundance is conveniently parametrized by
\begin{equation}
Y= {n\over s_{\rm SM}} \;,
\end{equation}
where $s_{\rm SM}$ is the SM entropy density. As the Universe expands, the latter evolves both due to the temperature change and the change in the 
effective number of the SM degrees of freedom (d.o.f.) in the plasma, which is particularly important at low temperatures. The entropy density and 
the Hubble rate can be parametrized as
\begin{equation}
s_{\rm SM}= {2\pi^2 \over 45} g_s T^3~,~ H = \sqrt{g_e \pi^2 \over 90 } \, {T^2 \over M_{\rm Pl}} \;,
\end{equation}
where $g_s$ is the SM d.o.f. number contributing to the entropy, $g_e$ is that contributing to the energy density, and $ M_{\rm Pl}$ is the Planck constant.
The SM entropy is conserved, $s_{\rm SM} a^3= {\rm const}$, with $a$ being the scale factor which allows us to trade the time variable 
for temperature $T$. For the Maxwell-Boltzmann distribution function, the reaction rate takes the form
\begin{equation}
\Gamma(\bar f f \rightarrow SS) = \langle  \sigma (\bar f f \rightarrow SS) v_r \rangle \, (n_f^{\rm eq})^2 \, , 
\end{equation}
where $\langle \dots \rangle$ stands for the thermal average, $\sigma$ is the reaction cross section which includes the symmetry factors 
for the initial and final states, $v_r$ is the relative velocity of the initial state particles and $n_f^{\rm eq}$ is the thermal equilibrium density of $f$. 
Then, introducing the variable
\begin{equation}
x\equiv {m_s \over T}\;,
\end{equation}
one can rewrite the Boltzmann equation as
\begin{equation}
{dY\over dx}  = 2 \; \sqrt{8\pi^2 M_{\rm Pl}^2 \over 45}\; {g_*^{1/2} m_s \over x^2 } \; \langle \sigma (\bar f f \rightarrow SS) v_r \rangle \, (Y_f^{\rm eq})^2 \;,
\label{dY/dx}
\end{equation}
where
\begin{equation}
g_*^{1/2}  = {g_s\over g_e^{1/2}} \;  \left( 1+ T {dg_s/dT \over 3g_s} \right)\;.
\end{equation}
The $g_*$ function of $T$ is known in the SM and shown in Fig.\,\ref{dof} \cite{Gondolo:1990dk}.

The above equation can be solved for $Y$ starting with a zero initial abundance. The observed value 
\begin{equation}
Y_\infty = 4.4 \times 10^{-10}\; {{\rm GeV}\over m_s} \,,
\label{Yinfty}
\end{equation}
then imposes a constraint on the input parameters of the model. The result, however, is sensitive to the SM bath temperature evolution before reheating, 
which exposes the dependence on the initial conditions inherent in all freeze-in models. In this work, we use the ``instant reheating'' approximation, 
that is, we assume that $T$ increases from 0 to $T_R$ abruptly and the DM production begins at  this point. The results, however, can be applied 
to a wider class of reheating models by replacing $T_R$ with the maximal temperature of the SM bath $T_{\rm max}$  and performing appropriate 
model-dependent rescaling $(T_{\rm max}/T_R)^\alpha$ of the coupling \cite{Cosme:2024ndc}.

The annihilation effects can trivially be included in the above equation by noting that, in kinetic equilibrium, $\Gamma(SS \rightarrow \bar f f) = 
\Gamma(\bar f f \rightarrow SS)  \times {(Y /Y^{\rm eq})^2} \,.$ The DM-SM system reaches kinetic equilibrium before the annihilation effect becomes 
important, hence it is safe to assume it for calculating back reaction.

\subsection{Simple estimates of the DM relic abundance}

It is instructive to estimate DM production $\bar f f \rightarrow SS$ in the perturbative regime when $g_s \simeq \,$const and $T \ll m_s$. In this case, 
the Boltzmann equation can be kept in the form
\begin{equation}
\dot n + 3Hn = 2 \, \Gamma (\bar f f \rightarrow SS) \;.
\label{BE}
\end{equation}
There are two regimes of interest, $m_f \ll m_s$ and $m_f \gg m_s$, which exhibit different qualitative features. In what follows, we consider them separately.
 
\subsubsection{$m_f \ll m_s$}

Using the Gelmini-Gondolo result \cite{Gondolo:1990dk}, the reaction rate for a single Dirac fermion $f$ in the regime $m_f < m_s $ can be written as 
\begin{eqnarray}
\Gamma (\bar f f  \rightarrow SS) &=&  2^2 \times  {1\over (2\pi)^6} \;     \int \sigma v_r \, e^{-E_1/T} e^{-E_2/T}  d^3 p_1 d^3 p_2  \nonumber \\
& =& 2^2 \times {2\pi^2 T\over (2\pi)^6}\;
\int_{4m_s^2}^\infty ds \; \sigma \;(s- 4 m_f^2) \sqrt{s} \, K_1 (\sqrt{s}/T) \;,
\label{G-G}
\end{eqnarray}
where $K_1(x)$ is the modified Bessel function of the first kind and we have explicitly factored out $2^2$ due to the spin d.o.f. of the initial fermions.
To compute the integral, we may use the large argument expansion $K_1(\sqrt{s}/T) \simeq \sqrt{\pi\over 2} \,T^{1/2} s^{-1/4} e^{-\sqrt{s}/T}$ and 
\begin{equation}
\sigma  (\bar f f  \rightarrow SS) = {\lambda^2_{hs} m_f^2 \over 64 \pi   m_h^4} \, {\sqrt{   1- {4m_s^2 \over s}}} \;.
\label{ffss}
\end{equation}
Due to the exponential fall-off of the integrand only values of $s$ around $4m_s^2$ contribute significantly. We may therefore approximate the ``slow'' 
function $s^{3/4} \rightarrow (4m_s^2)^{3/4}$ and drop the fermion mass, equation (\ref{G-G}), which leaves a non-trivial integral of the form 
\begin{equation}
\int_{4m_s^2}^\infty ds \;  (s- 4 m_s^2)^{1/2}  e^{-\sqrt{s}/T} \simeq (4 m_s T)^{3/2} e^{-2m_s /T} \times {1\over 2} \sqrt{\pi} \,,
\end{equation}
where we have reduced the integral to the Gamma function. This yields
\begin{equation}
\Gamma \simeq { \lambda^2_{hs} m_f^2 \,m_s^3\over 64 \pi^4   m_h^4} \, T^3 \,e^{-2m_s /T} \;.
\end{equation}
The Boltzmann equation (\ref{BE}) can then be solved analytically using ${d\over dt }(T^3 a^3)=0$ and setting the boundary condition $n(T_R)=0$.
The result is 
\begin{equation}
Y\simeq 0.2\times { \lambda^2_{hs} m_f^2 \,m_s^2 \,M_{\rm Pl}\over 64\pi^4  m_h^4} \, {e^{-2m_s /T_R} \over T_R }\;,
\label{Ylowmf}
\end{equation}
assuming a low temperature value $g_* \simeq 10$. For colored fermions, this result is to be multiplied by the color factor $N_c$, and other choices 
of $g_*$ can be accommodated using the scaling $Y \propto g_*^{-3/2}$.  The $T_R$ dependence is shared by other low temperature freeze-in 
models \cite{Arcadi:2024obp}.

The above result implies that the DM abundance is controlled predominantly by the ratio $m_s/T_R$. If this ratio is sufficiently large, ${\cal O}(10)$, 
the Higgs portal coupling $\lambda_{hs}$ can be large, up to order one. Therefore, one expects a significant decay width for the process $h \rightarrow SS$, 
\begin{equation}
\Gamma (h \rightarrow SS)= { \lambda^2_{hs} v^2 \over 32 \pi m_h} \sqrt{1- {4m_s^2 \over m_h^2}} \;.
\end{equation}
This both imposes a constraint on the model  \cite{Biekotter:2022ckj} and provides us with an avenue to probe the framework of freeze-in at stronger coupling.
The branching fraction of such ``invisible'' Higgs decay can be measured with percent-level accuracy at the LHC and subsequently at the FCC. The importance 
of this observable for low-$T$ freeze-in models was realized in Ref.\,\cite{Bringmann:2021sth}.

\subsubsection{$m_f \gg m_s$}

When the SM fermion is heavier than DM, the minimal value of $s$ is $4m_f^2$, hence,
\begin{eqnarray}
&& \Gamma (\bar f f  \rightarrow SS) = 2^2 \times {2\pi^2 T\over (2\pi)^6}\;
\int_{4m_f^2}^\infty ds \; \sigma \;(s- 4 m_f^2) \sqrt{s} \, K_1 (\sqrt{s}/T) \;,
\end{eqnarray}
while $\sigma$ remains as in equation (\ref{ffss}). The fermion mass cannot be neglected in the integrand such that the non-trivial part of the integral reduces to 
\begin{equation}
 \int_{4m_f^2}^\infty ds \;  (s- 4 m_f^2)^{3/2}  e^{-\sqrt{s}/T} \simeq (4 m_f T)^{5/2} e^{-2m_f /T} \times {3\over 4} \sqrt{\pi} \,.
\end{equation}
This yields
\begin{equation}
\Gamma \simeq {3   \lambda^2_{hs} m_f^4 \over 128 \pi^4   m_h^4} \, T^4 \,e^{-2m_f /T} \;.
\label{Gamma-heavyf}
\end{equation}
We see that the rate is {\it independent} of $m_s$ and suppressed by $e^{-2m_f /T}$ instead of the usual $e^{-2m_s /T}$. The resulting DM abundance is 
\begin{equation}
Y =  0.2 \times {3   \lambda^2_{hs} m_f^3 M_{\rm Pl} \over 128\pi^4   m_h^4}\, e^{-2m_f /T_R} \;,
\end{equation}
for $g_* \sim 10$. This result is independent of the DM mass and since the observational constraint (\ref{Yinfty}) is proportional to  $1/m_s$, 
the required Higgs portal coupling scales as
\begin{equation}
\lambda_{hs} \propto {1\over \sqrt{m_s}}\;.
\label{below-mu}
\end{equation}
It $decreases$ as DM becomes heavier. This is fundamentally different from the exponential scaling $\lambda_{hs} \propto e^{m_s/T_R}$ appearing 
in the regime $m_f \ll m_s$.

Generally, the DM production proceeds via annihilation of lighter fermions. Indeed, such a process is suppressed by  $e^{-2m_s/T}$, while for heavier 
fermions the suppression factor becomes $e^{-2m_f/T}$. However, the corresponding amplitude also involves a small Yukawa coupling. Therefore, there 
exists a range of DM masses, in which DM production is dominated by heavier fermions. We find that, in practice, this occurs in the region
\begin{equation}
m_e \lesssim m_s \lesssim m_\mu \;,
\end{equation} 
depending on the reheating temperature. The difference between the muon and electron Yukawa couplings is so large that it can compensate the exponential 
suppression. Let us estimate at which $T_R$ this occurs. The ratio of the two reaction rates at $T=T_R$  is given by
\begin{equation}
{\Gamma(\bar \mu  \mu \rightarrow SS) \over \Gamma (\bar e e  \rightarrow SS)  } \simeq  {3\over 2}\; {m_\mu^4 T_R \over m_e^2 m_s^3}  \,e^{-2(m_\mu-m_s) /T_R} \;.
\end{equation}
Focussing on the regime $m_s \gg T_R$, let us take $m_s = 50\,$ MeV. Then, one finds that the transition from the dominant electron to the dominant muon process 
occurs at $T_R \sim 10\;$MeV. At higher temperatures, the muon mode dominates.

It is important to note that DM produced in this regime is {\it relativistic} and can, in fact, be highly relativistic. This is in contrast to the case $m_f \ll m_s$.
Therefore, the non-relativistic DM approximation is not always adequate in this framework. In our numerical analysis, we take into account all the modes using 
relativistic formulation, thereby avoiding such problems.

\subsection{General case and hadronic contributions}
 
The above calculations apply when DM production is dominated by processes with elementary fermions in the initial state. At temperatures below the critical QCD temperature $T_c \sim 150\;$MeV, the quarks are no longer free and the reaction involves hadrons along with leptons. Our approach has therefore to be adjusted accordingly.

Let us first put equation (\ref{dY/dx}) in the form 
\begin{equation}
{dY\over dx}  = 2 \; \sqrt{8\pi^2 M_{\rm Pl}^2 \over 45}\; {g_*^{1/2} m_s \over x^2 } \;     
{\Gamma ( {\rm SM} \rightarrow SS)      \over s_{\rm SM}^2} \;,
\label{dY/dx-1}
\end{equation}
where $s_{\rm SM} $ is expressed as a function of $x$ and $\Gamma ( {\rm SM} \rightarrow SS)$ is the reaction rate per unit volume, 
with all the SM initial states included.

The production rate $\Gamma ( {\rm SM} \rightarrow SS)$ can be rewritten as the DM annihilation rate. Indeed, consider the reaction $1+2 \rightarrow 3+4$. 
Energy conservation implies that 
\begin{equation}
e^{-E_1/T} e^{-E_2/T}= e^{-E_3/T} e^{-E_4/T} \;.
\end{equation} 
Thus, for particles of the same spin, the momentum distribution functions satisfy $f_1 f_2 = f_3 f_4$. More generally, the initial and final state particles can have different
spins, in which case this relation generalizes to 
\begin{equation}
  f_1 f_2\,  | {\cal M}_{12\rightarrow 34}    |^2 = f_3 f_4  \,      | {\cal M}_{34\rightarrow 12}    |^2   \;,
\end{equation} 
where in $|{\cal M}_{ij\rightarrow kl}|^2$ one {\it averages over the initial and sums over the final spin states}, as usual. For example, the factor of 2 in $f_i$ for 
initial state fermions is compensated by the averaging factor 1/4 in the matrix element squared. Thus, one can replace the initial-state distribution functions 
with those of the final state, and equation (\ref{Gamma}) then makes it clear that 
 \begin{equation}
 \Gamma ( {\rm SM} \rightarrow SS) = \Gamma^{\rm th} ( SS \rightarrow {\rm SM}) \;,
 \end{equation}
where the superscript ``th'' implies that DM in the initial states is treated as {\it thermal}. Of course, this is a formal relation and DM produced via freeze-in is non-thermal.
Nevertheless, this expression facilitates the rate computation for general SM states.

Our next step is to factorize $ | {\cal M}(SS \rightarrow {\rm SM})|^2 $ using the fact that DM annihilation always proceeds via the Higgs mediator.  
We have 
\begin{equation}
 \vert {\cal M}_{SS\rightarrow {\rm SM}} \vert^2  = \vert {\cal M}_{SS\rightarrow h} \vert^2 \; 
 {1\over ( {\rm s}- m_h^2)^2}\; \vert {\cal M}_{h\rightarrow {\rm SM}} \vert^2  \;,
\label{factor}
\end{equation}
where the usual spin state summation/averaging is implied. The same relation applies to fermionic DM. The Higgs decay width is given by 
\begin{equation}
\Gamma_{h} =  {1\over 2 m_h} \;\int \left( \prod_{f} {d^3 {\bf p}_f \over (2 \pi)^3 2E_{f}} \right)~
\vert {\cal M}_{h\rightarrow {\rm SM }} \vert^2 ~ (2\pi)^4 \delta^4(p_h-\sum p_f)   \;,
\label{GammaHiggs}
\end{equation}
where $p_f$ are the momenta of the final state particles and a summation over all the SM final states is implied. 
This quantity is known in the SM for all Higgs masses, with variable accuracy. Putting these ingredients together in  equation (\ref{Gamma}), we find
\begin{eqnarray}
\Gamma^{\rm th}_{SS\rightarrow {\rm SM}} 
&=& \int \left( \prod_{i} {d^3 {\bf p}_i \over (2 \pi)^3 2E_{i}} \, f(p_i) \right)~\vert {\cal M}_{SS\rightarrow h} \vert^2      \; {1\over ( {\rm s}- m_h^2)^2} \times \left(2m_h \Gamma_h\right)\Bigl\vert_{m_h = \sqrt{\rm s} }\;.
\label{Gamma2-new}
\end{eqnarray}
Here, the SM Higgs mass in the last factor is $variable$ and given by the center-of-mass energy of the annihilating DM state, in analogy with 
the result of \cite{Burgess:2000yq}. This expression can be simplified further using the Gelmini-Gondolo formula (\ref{G-G}). Indeed, the integrand 
in (\ref{Gamma2-new}) has the form $\sigma v_r \, e^{-E_1/T} e^{-E_2/T}$. Since the relative (M\o ller) velocity of the initial particles is 
\begin{equation}
v_{r}= {1\over E_1 E_2} \; {1\over 2} \sqrt{ s} \sqrt{ { s} - 4m_s^2} \;,
\end{equation}
the role of $\sigma$ in (\ref{Gamma2-new}) is played by 
\begin{equation}
\sigma =  {1\over 2 \sqrt{ s} \sqrt{ { s} - 4m_s^2}} \; 
\vert {\cal M}_{SS\rightarrow h} \vert^2      \; {1\over ( {\rm s}- m_h^2)^2}\times  (2m_h \Gamma_h)\Bigl\vert_{m_h = \sqrt{\rm s} } \;.
\label{sig-new}
\end{equation}
We therefore conclude that at energies below the physical Higgs mass,
\begin{eqnarray}
\boxed{ \Gamma_{{\rm SM}\rightarrow SS} = \Gamma_{SS\rightarrow {\rm SM}}^{\rm th} =  {  T \over 2^5 \pi^4 m_h^4} \,
  \int_{4m_s^2}^\infty ds\;  \sqrt{s(s-4m_s^2)}\, K_1 \left(\sqrt{s}/T\right)  \,  \Gamma_h\left(m_h = \sqrt{s}\right)   \,      \left\vert {\cal M}_{SS\rightarrow h} \right\vert^2          } 
 \label{Gamma4-new}
\end{eqnarray}
In this form, the result can easily be converted to the fermionic DM case by multiplying the right hand side by the spin d.o.f. factor $2^2$. 
This expression does not assume that DM is non-relativistic. Indeed, the assumption of the Maxwell-Boltzmann distribution applies at $E/T \gg 1$, which is 
satisfied in our case either due to DM  or the {\it initial state SM particles} being non-relativistic ($m_f > T$).  When the non-relativistic DM limit applies, the integral is
dominated by the vicinity of $s= 4 m_s^2$ and can be simplified further. In particular, it matches our results for $ \Gamma_{\bar f f \rightarrow SS}$ presented earlier.
Note that in our convention, $$\vert{\cal M}_{SS\rightarrow h} \vert^2 = 1/2 \times \lambda_{hs}^2 v^2 \;,$$ since the initial state symmetry factor is included in the amplitude.

\begin{figure}[h!]
    \centering
    \includegraphics[width=0.45 \textwidth]{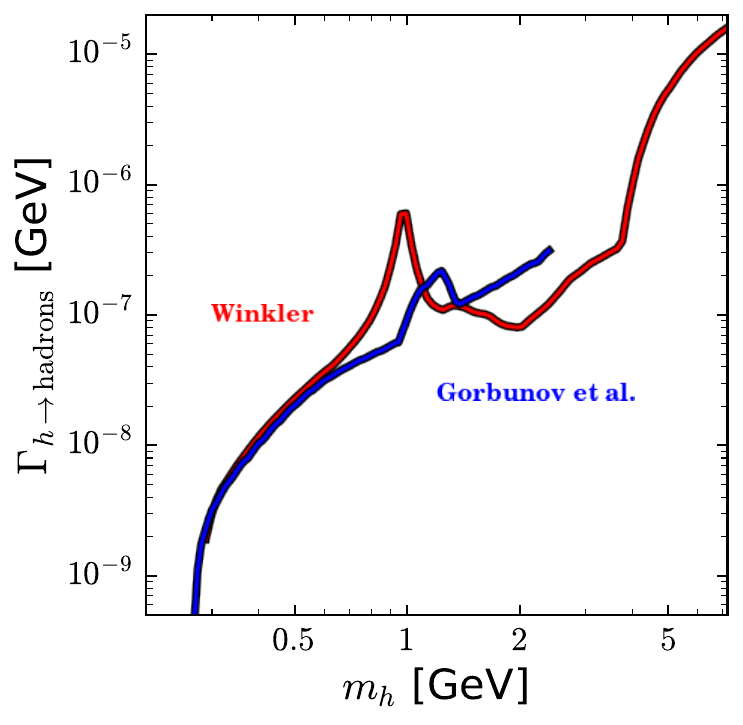}
    \caption{ Decay width of $h\rightarrow \;$hadrons for variable Higgs mass $m_h$ 
    according to Refs.\,\cite{Winkler:2018qyg} and \cite{Gorbunov:2023lga}.}
    \label{G-hadron}
\end{figure}

The SM Higgs decay width $\Gamma_h (m_h)$ for all relevant Higgs masses $m_h$ has been computed by different groups. The most challenging
calculation is $\Gamma(h\rightarrow \,$hadrons) for GeV and sub-GeV Higgs masses. In this mass range, the processes are non-perturbative requiring a more sophisticated 
approach. Below $m_h \sim\,$0.5 GeV, only the two pion mode is allowed which can be treated in the chiral perturbation theory and using the low energy theorems
\cite{Vainshtein:1980ea}, \cite{Voloshin:1986hp}, \cite{Dawson:1989yh}, while at higher masses other approaches including dispersive relations, etc. can be employed.
In our numerical analysis, we use two recent estimates  by Winkler \cite{Winkler:2018qyg}  and Gorbunov {\it et al.} \cite{Gorbunov:2023lga}, shown in Fig.\,\ref{G-hadron}.
While their results agree well up to $m_h \sim 0.5\;$GeV, at higher Higgs masses the discrepancy grows to up to an order of magnitude\footnote{The peaks seen 
in Fig.\,\ref{G-hadron} are associated with resonances, e.g.~$f_0(980)$, in the $\pi\pi$ scattering, although the impact of such resonances is different in the two
approaches.}. The consequent difference in the DM abundance gives us a reasonable estimate of the hadronic uncertainty in our analysis. Ref.\,\cite{Winkler:2018qyg}
matches its result to the perturbative calculation at the $\bar c c $ threshold, while the analysis of \cite{Gorbunov:2023lga} stops at $m_h$ below 3 GeV. 
In our computation, we take the two approaches to coincide above 3 GeV.

We use the input from Fig.\,\ref{G-hadron} above $m_h \simeq 2m_\pi$ to $m_h \simeq 2m_c$, while treating the decay $h \rightarrow \;$quarks perturbatively 
at higher masses. After including the lepton and photon decay modes, we solve the Boltzmann equation (\ref{dY/dx-1}) numerically. The photon coupling to the Higgs 
boson is generated via loops of the SM particles, and the corresponding decay width is given by \cite{Spira:1995rr}
\begin{equation}
\label{eq:Sdecwidthphph}
 \Gamma(h \to \gamma \gamma) = \frac{G_F \alpha_{\rm em}^2 m_h^3 }{128 \sqrt{2} \pi^3} \Bigg| \sum_f N^f_c Q_f^2 A_f(\tau_f) + N^W_c  Q_W^2 A_W(\tau_W)  \Bigg|^2,
\end{equation}
where $\alpha_{\rm em}$ is the fine structure constant the sum runs over fermions $f$ and $W$ inside the loop. In this expression, $N_c$ is the number of colors of the species at hand, 
$Q_i$ is the charge and $G_F$ is the Fermi coupling constant. The result is parametrized in terms of 
$ \tau_{\rm x} = \frac{m_h^2}{4 m_{\rm x}^2} \, , $ with the loop functions 
\begin{align}
 A_f(\tau) &= 2 (\tau + (\tau -1) f(\tau))/\tau^2 \,, \\
 A_W(\tau) &= -(2 \tau^2  + 3\tau  + 3 (2 \tau -1) f(\tau) )/\tau^2 \,,
\end{align}
and 
\begin{equation}
  f(\tau) = \begin{cases}
    \text{arcsin}^2 \sqrt{\tau} \hspace{1cm} &\text{for} \,\, \tau \le 1 \,, \\
    -\frac{1}{4}\left(\log \frac{1+\sqrt{1-\tau^{-1}}}{1-\sqrt{1-\tau^{-1}}} -i\pi \right)^2 &\text{for} \,\, \tau > 1 \,.
  \end{cases}
\end{equation}
In practice, we find that the photon contribution is unimportant in the entire mass range of interest. The analogous gluon contribution is also small above 
the charm threshold \cite{Winkler:2018qyg}, while below the threshold it is included in the hadronic mode. 
 \begin{figure}[h!]
    \centering
    \includegraphics[width=0.9 \textwidth]{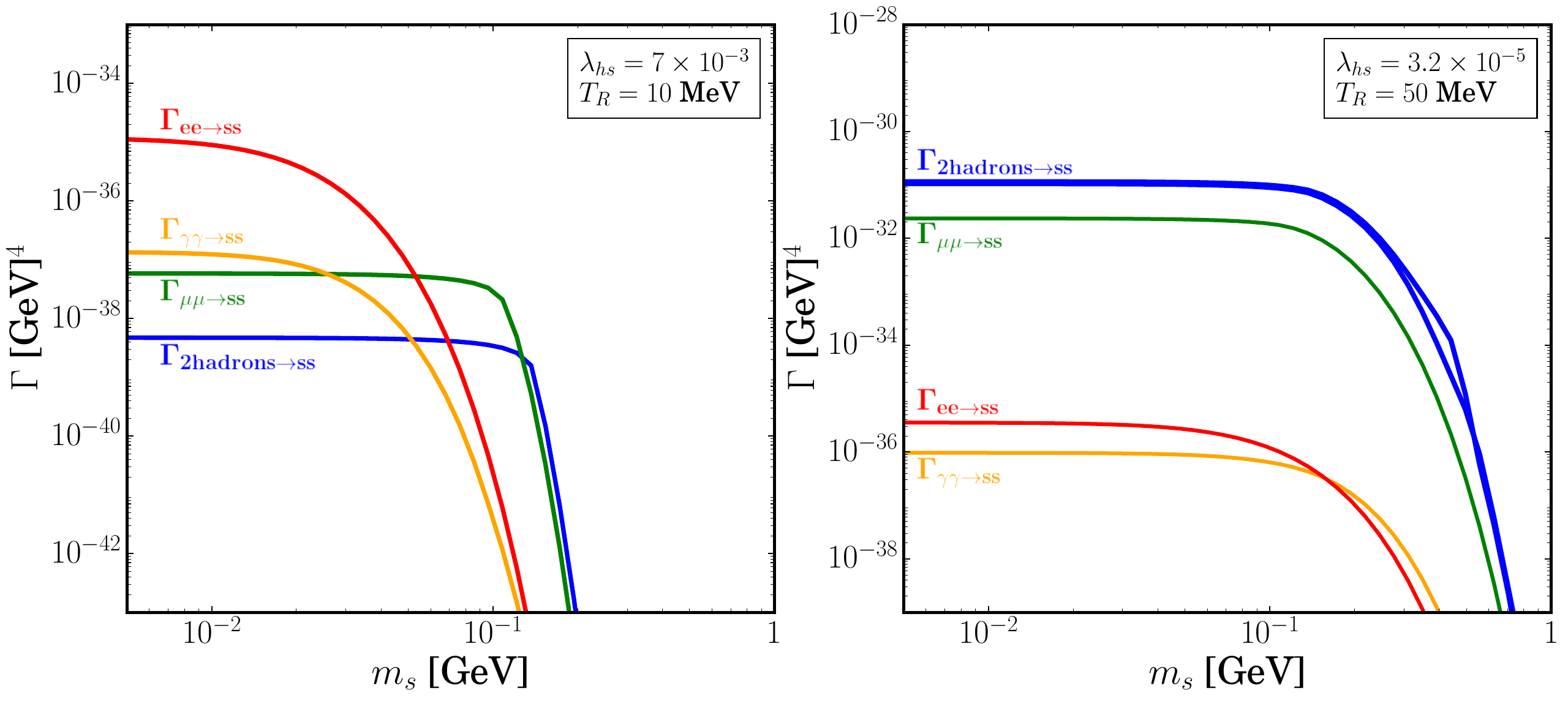}
    \caption{Individual reaction rates. The broadening of the blue curve in the right panel represents hadronic uncertainties.
    }
    \label{rates-scalar}
\end{figure}

The relative importance of the different channels for sub-GeV $m_s$ is shown in Fig.\,\ref{rates-scalar}. At higher temperatures, the hadronic mode 
dominates due to the stronger coupling to the Higgs boson, while at low temperatures, the electron mode becomes more important. We observe 
an exponential suppression of the rates at high DM masses, while at low masses the rate profile becomes flat. For the hadronic and muon modes, 
the flat patch occurs in the regime $m_\pi, m_\mu \gg m_s$,  where our result (\ref{Gamma-heavyf}) applies. For the electron and photon modes, 
the temperature is higher than all the masses involved which makes the reaction rate mass-independent. In this limit ($T \gg m_e, m_s$), 
the Maxwell-Boltzmann approximation receives significant corrections, yet it still gives a reasonable estimate for the reaction rate, typically 
within a factor of two \cite{Lebedev:2019ton}, \cite{Lebedev:2023uzp}. This is inconsequential in most of the parameter space, where the muon 
and hadronic modes dominate. In the region $10\,{\rm MeV}\gtrsim T_R \gg m_s$, however, the leading process is $\bar e e \rightarrow SS$ 
and the above (Pauli-blocking) correction applies. In practice, the effect of this correction on the parameter space analysis is modest: it amounts 
at most to  $\sim 40\%$ increase in $\lambda_{hs}$ within the above small region.

\subsection{Parameter space analysis}

\begin{figure}[h!]
    \centering
    \includegraphics[width=0.49 \textwidth]{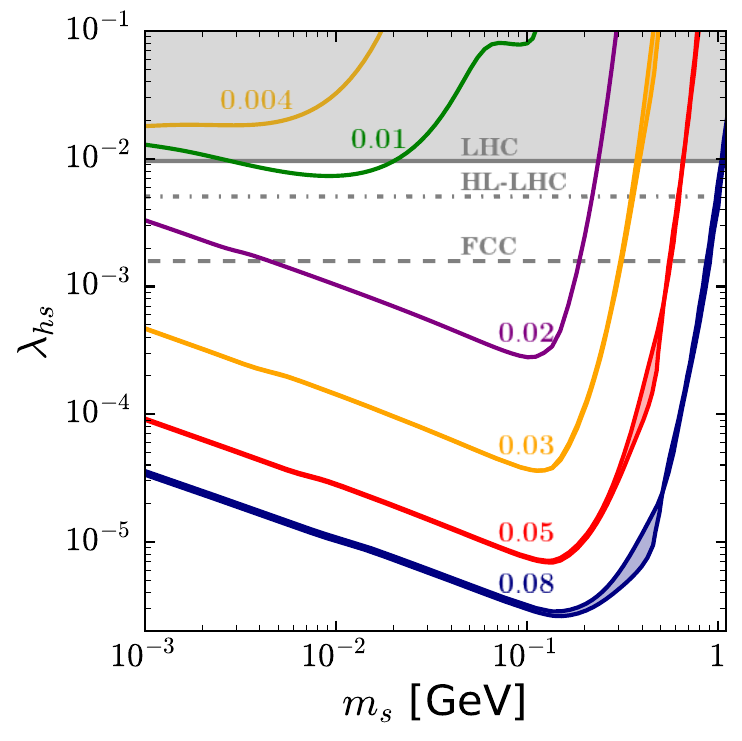}
    \includegraphics[width=0.49 \textwidth]{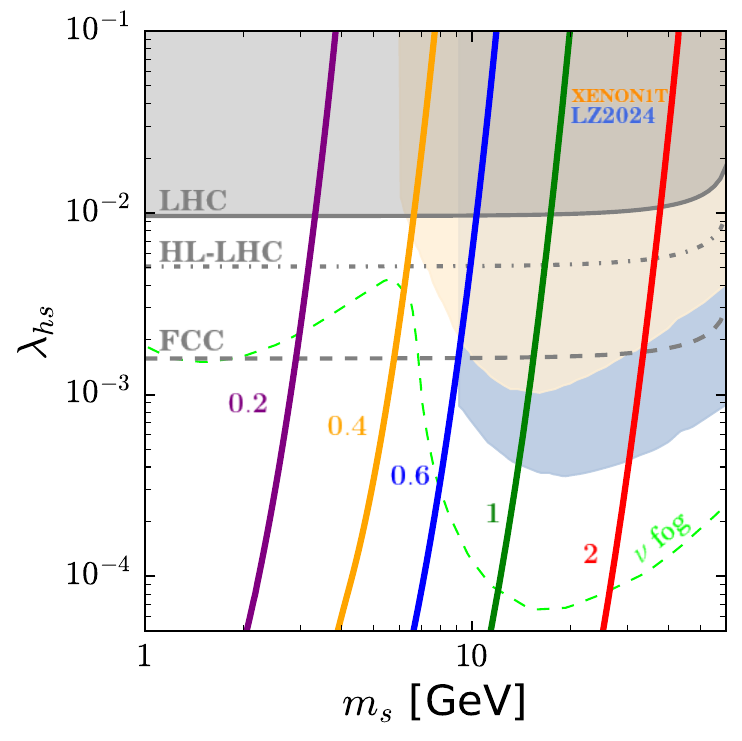}
    \caption{Parameter space for scalar DM freeze-in. Along the colored curves marked by $T_R$ in GeV, 
    the correct relic density is reproduced. The shaded areas are excluded by the LHC and direct DM detection bounds.
    Sensitivities of the HL-LHC and FCC are shown by the grey dashed lines, while the neutrino ``fog'' for direct DM detection \cite{Billard:2021uyg} 
    is represented by the green dashed line.}
    \label{par-space-1}
\end{figure}

Our main results are presented in Fig.\,\ref{par-space-1}. It shows the allowed parameter space in terms of $\lambda_{hs}$ and $m_s$.
The colored curves represent the correct relic density for a given reheating temperature $T_R$. For the DM mass above the muon threshold, 
the Higgs portal coupling exhibits the usual exponential growth with $m_s$ (see equation (\ref{Ylowmf})), while below the threshold its behavior 
is described by (\ref{below-mu}). This occurs because of the hierarchy in the muon vs electron couplings to the Higgs. Even though the muon 
abundance is exponentially suppressed at $T_R < m_\mu$, the reaction $\bar \mu \mu \rightarrow SS$ is more efficient than its electron analogue 
due to the larger Yukawa coupling. This feature starts fading away at very low temperatures, $T_R < 10\;$MeV.

The shaded grey area is excluded by the LHC search for invisible Higgs decay (BR$_{\rm inv}\lesssim 10\,$\% \cite{ATLAS:2023tkt}), while 
the dashed and dash-dotted lines represent sensitivities of the FCC and HL-LHC  to this mode. We take the HL-LHC benchmark goal to be 
BR$_{\rm inv}=3\,$\% \cite{RivadeneiraBracho:2022sph}, while that of the FCC to be BR$_{\rm inv}=0.3\,$\% \cite{FCC}. Fig.\,\ref{par-space-1} 
shows that $T_R \gtrsim10\,$MeV is compatible with all the constraints in a range of DM masses. The corresponding parameter space can be 
probed by the upcoming experiments.

The hadronic uncertainties do not impact these results significantly. We observe broadening of the relic abundance curves at higher $T_R \gtrsim 50\,$MeV 
(but below $T_c$), in which case the pion abundance is not suppressed significantly and hadrons make a tangible contribution to DM production. 
The two approaches to $h\rightarrow \;$hadrons transition lead to very similar results in most of the parameter space and the largest discrepancy 
amounts to  about 50-100\% correction to $\lambda_{hs}$ in a narrow range of DM masses around 0.5 GeV. This uncertainty is, however, specific 
to the parameter space regions which can hardly be probed at colliders and thus makes a little impact on our predictions. 

It is important to remember that we are using a {\it dilute hadron gas approximation}, which is adequate away from the QCD critical temperature. 
Above $T_c$, we approximate the system by a fermion gas, while below $T_c$, we have a hadron gas with Boltzmann-suppressed meson abundance, 
which can therefore be treated in a dilute gas approximation too. At $T_R\sim T_c$, these approaches fail, which prompts us to avoid this region. 
We note that if the system starts at $T_R>T_c$, it will go through the confinement phase. However, it does not pose a problem since DM production 
is dominated by the highest temperatures. To avoid proximity of $T_R$ and $T_c$, in Fig.\,\ref{par-space-1} we restrict ourselves to the $T_R$ range 
below 80 MeV and above 200 MeV. 

The DM direct detection (DD) experiments \cite{LZCollaboration:2024lux,XENON:2018voc} impose an additional constraint on the model. 
The nucleon-DM scattering cross section in our model is given by
 \begin{equation}
     \sigma_{sN} \simeq \frac{\lambda_{hs}^2 f_N^2}{4 \pi} \frac{m_N^4}{m_h^4 m_s^2}\,,
 \end{equation}
 where $f_N \simeq 0.3$ and $m_N \simeq 1$ GeV. The strictest constraint for $m_s \gtrsim 9\;$GeV is imposed 
by the recent LZ2024 result \cite{LZCollaboration:2024lux}.

The right panel of Fig.\,\ref{par-space-1} shows that the DM DD constraints supersede those from the Higgs invisible decay 
at $m_s \gtrsim 6\,$GeV \cite{Arcadi:2024wwg}. This defines the upper mass bound for DM searches via Higgs decay. On the other hand, 
the lower mass bound is imposed by cosmological considerations only, e.g.~by structure formation constraints. 

Finally, let us note that DM in the parameter range of Fig.\,\ref{par-space-1} is {\it cold}. In the Boltzmann-suppressed regime $T_R \ll m_s$, 
this is automatic. For $T_R \gg m_s$, DM becomes non-relativistic above $T \sim 0.1 \;$MeV. This can be seen as follows. At $T_R \lesssim 10\;$MeV, 
DM is produced via electron annihilation. Its energy is determined by the visible sector's (SM bath) temperature and it scales as such with time. 
Since $m_s \gtrsim \;$MeV, it becomes non-relativistic at and below MeV temperatures. At higher reheating temperatures, $T_R \gtrsim 10\;$MeV, 
the dominant production modes are given by the pion and muon annihilation processes. Hence, the initial DM energy is determined by the pion/muon 
mass of around 100 MeV. It scales down to around an MeV, which corresponds to its mass, when the temperature drops by a factor of hundred. Therefore, around $T\gtrsim 0.1\;$MeV, 
DM is non-relativistic. We thus conclude that, at the stage of structure formation corresponding to keV-scale temperatures, DM is already cold.

\subsection{Non-thermalization constraint}

The crucial feature of freeze-in DM is that it never reaches thermal equilibrium with the SM bath. This is ensured by 
\begin{equation}
\Gamma ({\rm SM}\rightarrow SS) \not=  \Gamma (SS \rightarrow  {\rm SM}) \;,
\end{equation}
such that the detailed balance is not maintained.  As a result, the DM density differs from its equilibrium value at the SM bath temperature $T$.
 \begin{figure}[t!]
    \centering
    \includegraphics[width=0.9 \textwidth]{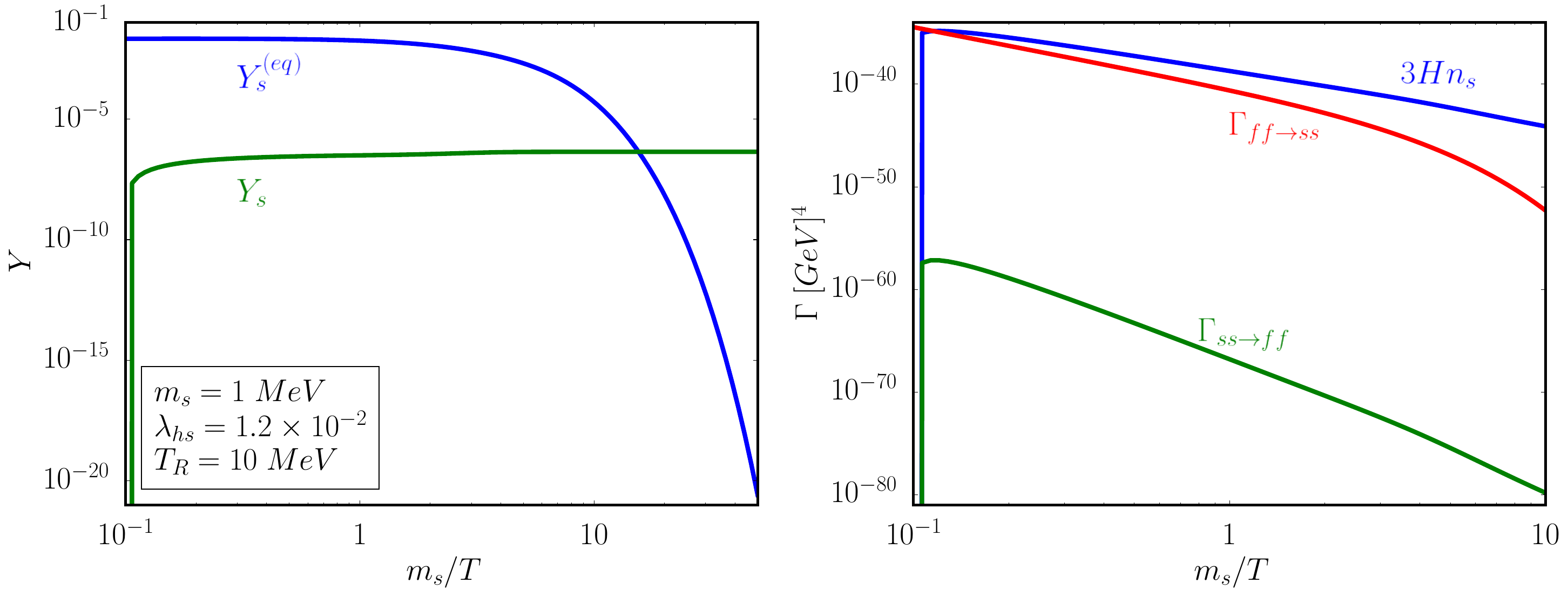}
    \caption{ {\it Left:} Evolution of the scalar DM abundance $Y_s$ vs its thermal equilibrium value $Y_s^{\rm (eq)}$. 
    {\it Right:} Reaction rates $\bar ff \leftrightarrow SS$ vs expansion rate $3Hn_s$. }
    \label{therm-1}
\end{figure}

In the regime $T\ll m_s$,  this is trivially verified and thermal equilibrium requires a rather large coupling (see e.g. \cite{Cosme:2023xpa}).
In our current work, we also consider the regime $T\gtrsim m_s$, which applies, in particular, at $m_s < m_\mu$.  In this case, the result is not obvious 
and warrants proper investigation. Fig.\,\ref{therm-1} shows a representative example of $T_R =10\;$MeV and $m_s =1\;$MeV at $\lambda_{hs}=0.012$. 
Even though the Higgs portal coupling is substantial, the light SM fermions' coupling to the Higgs boson is very small, which makes the reactions inefficient. 
We observe that the DM density is orders of magnitude below its equilibrium value in the relativistic regime. This can be understood by comparing 
the different terms in the Boltzmann equation. The right panel of Fig.\,\ref{therm-1} shows that the DM production rate is comparable to the expansion rate  $3Hn$ 
only for a brief period of time. Most of the DM quanta are produced at this stage. The DM annihilation rate, on the other hand, is many orders of magnitude 
below the expansion rate, which shows that equilibrium is never achieved\footnote{For the parameters of Fig.\,\ref{therm-1}, thermalization is achieved 
at $T_R \sim 200\;$MeV, in which case the quark and lepton modes are activated.}. Similarly, we have verified the non-thermalization condition in the entire 
parameter space of interest.

In all of the above considerations, we have assumed ``instant reheating'', i.e.~that the temperature increases abruptly from 0 to $T_R$. This approximation applies, 
first and foremost, to a class of reheating models with a flat temperature profile \cite{Cosme:2024ndc}. Even though the SM sector temperature stays constant for an
extended period of time before reheating, the produced DM gets diluted such that the result is dominated by the latest moments with the same $T$\footnote{For
$T_R>m_s$, corresponding to the standard freeze-in scenario, DM production is also dominated by late times, $T\sim m_s$. Similarly, DM thermalization is 
most efficient at the lowest temperature compatible with the relativistic regime \cite{Lebedev:2023uzp,Lebedev:2021ixj}. See also a discussion in \cite{Bernal:2024yhu}.}.
One finds that the resulting $\lambda_{hs}(m_s)$  in the UV-complete reheating model matches closely to that obtained in the instant reheating approximation. 
More generally, our results apply to a larger class of models with $T_{\rm max} > T_R$, as long as DM remains non-thermal. The DM production is dominated 
by the highest temperature $T_{\rm max}$ such that, in order to obtain the corresponding abundance, one replaces $T_R \rightarrow T_{\rm max}$ in our analysis 
and accounts for entropy production before reheating by introducing a factor $(T_R/T_{\rm max})^\alpha$, with a model-dependent $\alpha$, into $Y$ 
\cite{Cosme:2024ndc}. Therefore, our results can be applied to a class of reheating models, in which the maximal temperature is not high enough to thermalize DM.

\section{Fermion dark matter}
\label{Sect:fermionDM}

Let us now perform an analogous analysis for Majorana fermion DM $\chi$. The main difference from the scalar case lies in the energy dependence 
of the DM production cross section, which leads to certain distinct features. The CP-even and CP-odd Higgs-fermion couplings exhibit somewhat different 
phenomenology \cite{Lopez-Honorez:2012tov}, so we will consider these two cases separately.

\subsection{CP-even coupling}   

The low energy coupling is given by ${m_f\over \Lambda m_h^2 } \, \bar f f \, \bar \chi \chi $. As before, let us first consider the simpler case of DM production 
by elementary SM fermions at $T \ll m_\chi$. For light fermions, $m_f \ll m_\chi$, the $\bar ff \rightarrow  \chi \chi$ cross section is
\begin{equation}
\sigma (\bar ff \rightarrow  \chi \chi)= {m_f^2 \over 8 \pi \Lambda^2 m_h^4 s} \, (s- 4m_\chi^2)^{3/2} \, (s-4 m_f^2)^{1/2} \;,
\end{equation}
where $s$ is the Mandelstam variable and $\sigma$ includes proper symmetry factors for Majorana fermions. The result exhibits the usual velocity 
suppression characteristic of fermion production via a CP-even coupling.

Computing the reaction rate at $T \ll m_\chi$ in the usual way, we get
\begin{equation}
\Gamma (\bar ff \rightarrow  \chi \chi)= {3 m_f^2  \over 4\pi^2 \Lambda^2  m_h^4} \, m_\chi^4 T^4 e^{-2 m_\chi /T}\;,
\end{equation}
where we suppress the color multiplicity factor $N_c$ for quarks. The temperature dependence $T^4 e^{-2 m_\chi /T}$ is steeper than that in the scalar DM case 
due to the velocity suppression of the cross section. The resulting DM abundance in the freeze-in limit is found by integrating $\dot n + 3Hn = 2 \Gamma$, 
yielding for a single Dirac fermion $f$,
\begin{equation}
Y= {3 m_f^2  \over 4\pi^2 \Lambda^2  m_h^4 \, c \tilde c} \, m_\chi^3 \,e^{-2 m_\chi /T_R}\;,
\end{equation}
with $c= \sqrt{\pi^2 g_* \over 90} M_{\rm Pl}^{-1} ~,~ \tilde c = {2\pi^2 g_* \over 45}$. Above the muon threshold, one finds $m_\chi \sim 10 \,T_R$ for typical 
parameter values and a TeV scale $\Lambda$. As before, the abundance scales as $g_*^{-3/2}$ with the number of the SM d.o.f. 

All the production modes are included via an analogue of  equation (\ref{Gamma4-new}). Its right hand side should be multiplied by $2^2$ to account for 
the spin d.o.f.~of $\chi$, while the square of the matrix element averaged over the initial spin (and including the symmetry factor 1/2 for the initial state) 
becomes
\begin{equation}
 \left\vert {\cal M}_{\chi \chi \rightarrow h} \right\vert^2 = {2 v^2 \over \Lambda^2}\, (s -4 m_\chi^2)~.
\end{equation}
This yields for $s\ll m_h^2$,  
\begin{eqnarray}
\Gamma_{ {\rm SM}\rightarrow \chi\chi} = \Gamma_{\chi\chi\rightarrow {\rm SM}}^{\rm th} =  {  v^2 T \over 8 \pi^4 m_h^4 \Lambda^2} \times
\int_{4m_\chi^2}^\infty ds\;  \sqrt{s}\, ({s-4m_\chi^2})^{3/2}\, K_1 (\sqrt{s}/T)  \times \Gamma_h (m_h = \sqrt{s}) \;.
\label{Gamma-f10}
\end{eqnarray}
We observe a steeper $s$-dependence of the integrand compared to the scalar case.

The scale $\Lambda$ is constrained by the invisible Higgs decay. The decay width into Majorana fermions is given by
\begin{equation}
\Gamma (h \rightarrow \chi \chi) ={m_h \over 4\pi}\, {v^2\over \Lambda^2}\, \left(1-{4m_\chi^2 \over m_h^2}\right)^{3/2} \;.
\end{equation}
Using the benchmark values for ${\rm BR_{inv}}(h \rightarrow {\rm DM})$ as in the scalar case, we obtain the LHC constraint 
as well as the future collider sensitivity to $\Lambda$.
\begin{figure}[t!]
    \centering
    \includegraphics[width=0.49 \textwidth]{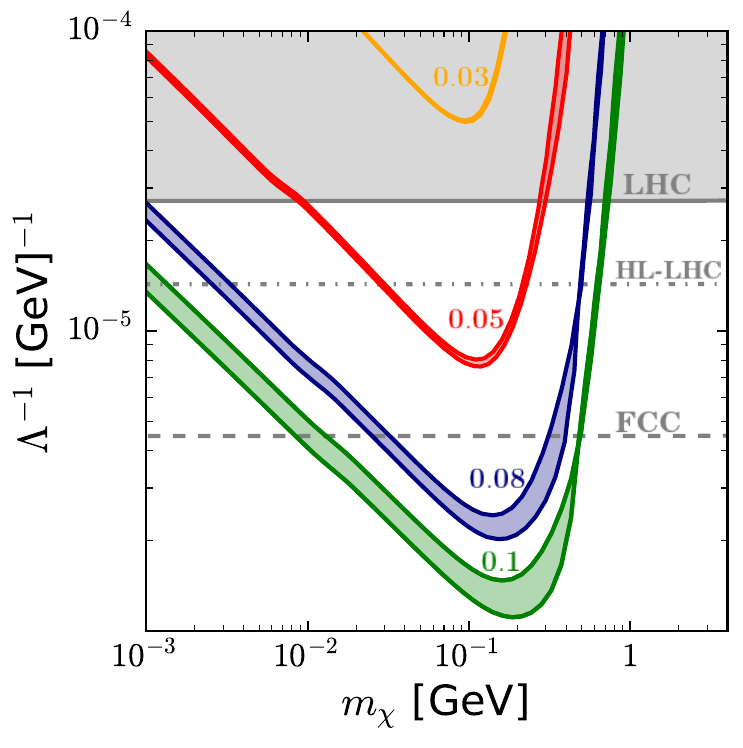}
    \includegraphics[width=0.49 \textwidth]{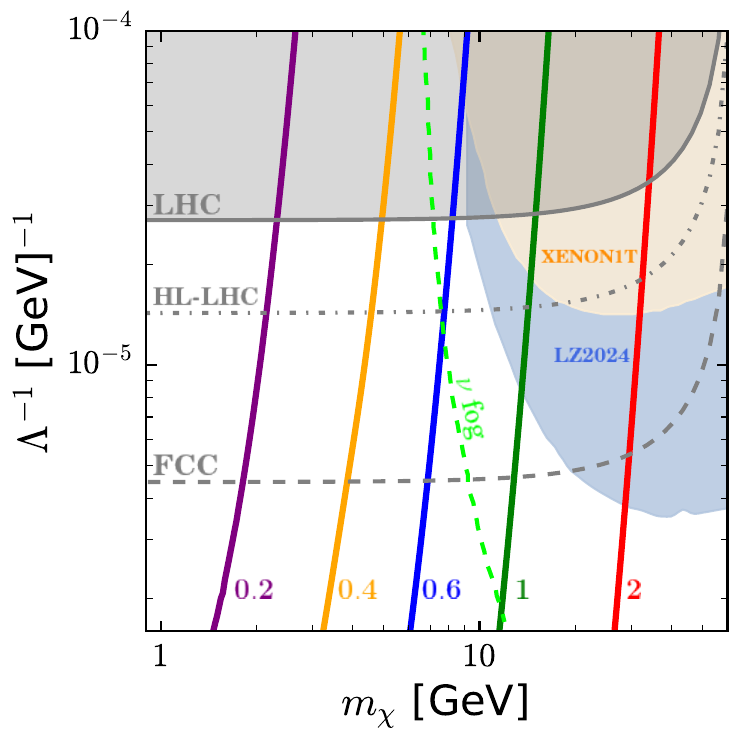}
    \caption{Parameter space for fermion DM freeze-in. Along the colored curves marked by $T_R$ in GeV, 
    the correct relic density is reproduced. The shaded areas are excluded by the LHC and direct DM detection bounds.
    Sensitivities of the HL-LHC and FCC are shown by the grey dashed lines, while the neutrino ``fog'' 
    for direct DM detection \cite{Billard:2021uyg} is represented by the green dashed line.}
    \label{par-space-2}
\end{figure}

Combining the constraints and including all the relevant production channels, we find the allowed parameter space in Fig.\,\ref{par-space-2}. 
The reheating temperatures above 40 MeV are compatible with the correct DM relic abundance. This value is higher than the minimal $T_R\sim 10\;$MeV 
allowed in the scalar case. We also observe that the hadronic uncertainty in the relic abundance leads to the band broadening from $m_\chi \sim 0.5\;$GeV 
down to the MeV masses. This feature distinguishes the fermion and scalar DM cases. For fermions, the integrand in (\ref{Gamma-f10}) has a steeper 
$s$-dependence which makes the result sensitive to higher ``virtual Higgs'' masses and hence to the corresponding uncertainties in $\Gamma_h$, 
even for very light DM. In the scalar case, the integral is dominated by low $s$ such that at $m_s \ll m_\pi$ the uncertainties are small. The size of 
the ``error bars'' is, nevertheless, modest for fermions: it does not exceed $\sim $50\% for the entire DM mass range (excluding $T_R\sim T_c $ 
between 100 and 200 MeV).

The dominant production modes remain the same as in the scalar case (Fig.\,\ref{rates-scalar}). Since very low reheating temperatures are incompatible 
with the Higgs data, light DM is produced via the muon and pion annihilation. This is represented in Fig.\,\ref{par-space-2} by the upward coupling trend 
as $\chi$ gets lighter. Above $T_R \sim 200\;$ MeV, the production modes involve quarks and leptons. Similarly to the scalar case, DM is cold as 
long as $m_\chi \gtrsim \;$MeV.

Above $m_\chi \sim 6\;$GeV, the direct detection bounds become important, with the spin-independent DM-nucleon cross section \cite{Arcadi:2024wwg} 
 \begin{equation}
     \sigma_{\chi N} = \frac{4}{\pi \Lambda^2 m_h^4} \frac{m_N^4 m_\chi^2}{(m_N + m_\chi)^2}f_N^2\,. 
 \end{equation}
Below $6$ GeV, the Higgs invisible decay imposes the only constraint and also presents a venue to probe the model. As in the scalar case, 
we find a wide range of DM masses with good detection prospects.

\subsection{CP-odd coupling}

The corresponding DM-SM coupling is ${m_f\over \Lambda_5 m_h^2 } \, \bar f f \, \bar \chi \, i\gamma_5 \, \chi $. The $\bar ff \rightarrow  \chi \chi$ cross section 
at low energy becomes
\begin{equation}
\sigma (\bar ff \rightarrow  \chi \chi)= {m_f^2 \over 8 \pi \Lambda^2_5 m_h^4 } \, (s- 4m_\chi^2)^{1/2} \, s^{1/2} \;.
\end{equation}
We observe that the strong velocity suppression seen in the CP-even case disappears. The reaction becomes more efficient, and in the $m_f \ll m_\chi$ limit 
we have
\begin{equation}
\Gamma (\bar ff \rightarrow  \chi \chi)= { m_f^2  \over 2\pi^2 \Lambda^2_5  m_h^4} \, m_\chi^5 T^3 e^{-2 m_\chi /T}\;.
\end{equation}
This result is similar to the scalar production rate considered earlier. It is enhanced by the factor $2/3 \times m_\chi /T$ compared to the CP-conserving 
fermion case. The corresponding DM abundance generated by a single Dirac fermion is found to be 
\begin{equation}
Y= { m_f^2  \over 2\pi^2 \Lambda^2_5  m_h^4 \, c \tilde c} \, {m_\chi^4 \over T_R}\, e^{-2 m_\chi /T_R}\;.
\end{equation}
Due to the exponential dependence on $m_\chi /T_R$, the numerical results for the CP-odd and CP-even cases do not differ significantly.
\begin{figure}[t!]
    \centering
    \includegraphics[width=0.49 \textwidth]{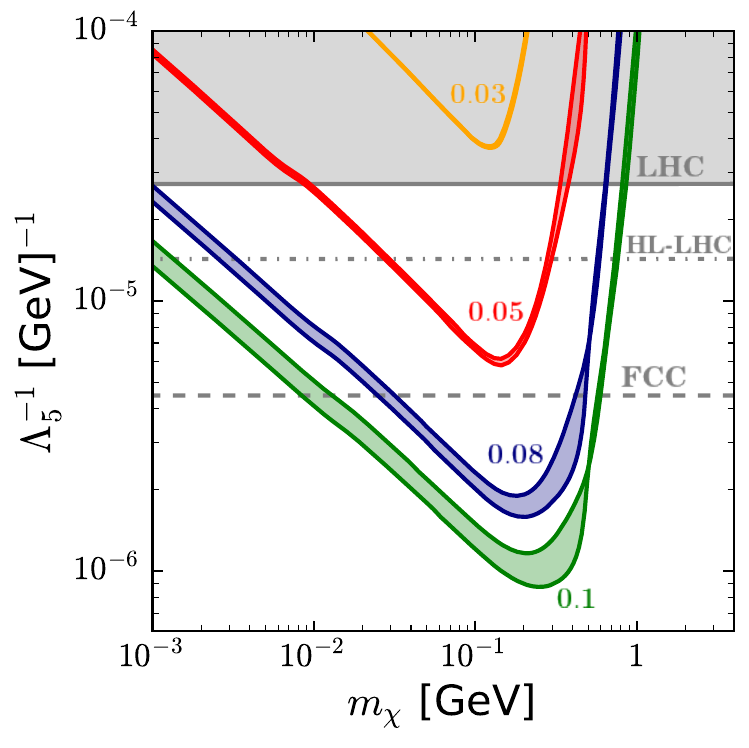}
    \includegraphics[width=0.49 \textwidth]{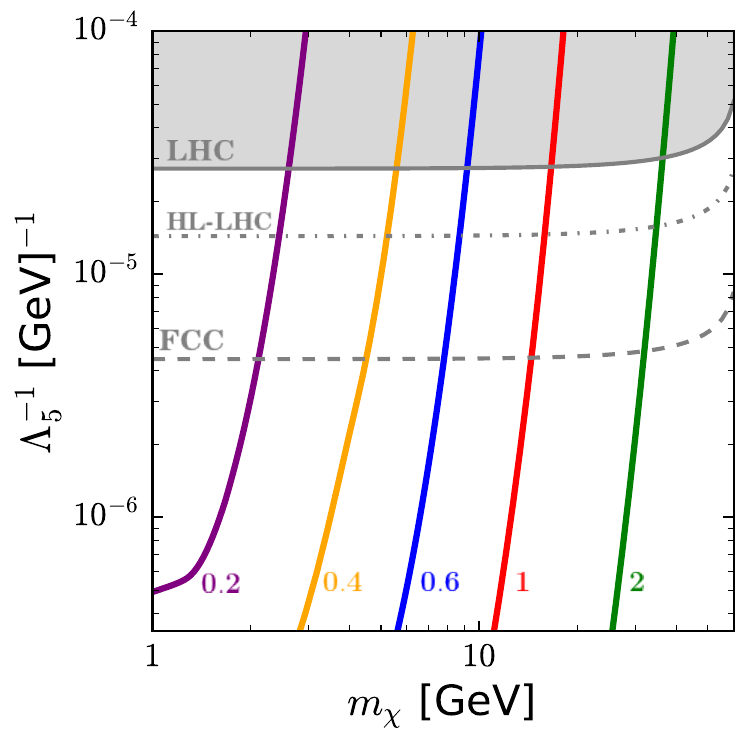}
    \caption{Parameter space for CP-odd fermion DM freeze-in. Along the colored curves marked by $T_R$ in GeV, the correct relic density is reproduced. 
    The shaded area is excluded by the LHC. Sensitivities of the HL-LHC and FCC are shown by the grey dashed lines.}
    \label{par-space-3}
\end{figure}

The analogue of equation (\ref{Gamma4-new}) is obtained with the help of  
\begin{equation}
\left\vert {\cal M}_{\chi \chi \rightarrow h}^{\gamma_5} \right\vert^2 = {2 v^2 \over \Lambda_5^2}\, s \;,
\end{equation}
and reads 
\begin{eqnarray}
 \Gamma_{{\rm SM}\rightarrow \chi\chi}^{\gamma_5} = \Gamma_{\chi\chi\rightarrow {\rm SM}}^{\rm th~ \gamma_5} =  
 \frac{v^2 T}{8 \pi^4 m_h^4 \Lambda^2_5}   \times \int_{4m_\chi^2}^\infty ds\;  {s}^{3/2}\, \sqrt{s-4m_\chi^2}\, 
 K_1 (\sqrt{s}/T)  \times \Gamma_h (m_h = \sqrt{s}) \;.
 \label{Gamma-f2}
\end{eqnarray}
The Higgs invisible decay width becomes
\begin{equation}
\Gamma (h \rightarrow \chi \chi) ={m_h \over 4\pi}\, {v^2\over \Lambda^2_5}\, \left(1-{4m_\chi^2 \over m_h^2}\right)^{1/2} \;.
\end{equation}

The corresponding parameter space is shown in Fig.\,\ref{par-space-3}. We observe that most features are shared by the CP-even and CP-odd cases, 
except the scale $\Lambda_5$ is allowed to be somewhat higher than $\Lambda$ due to the absence of the strong velocity suppression in production 
of non-relativistic DM. The main difference is that the direct DM detection bounds are very loose in the CP-odd case. This is because of the large 
velocity suppression ($10^{-6}$) of the detection rate, which makes the invisible Higgs decay the only venue to probe the model \cite{Arcadi:2024wwg}.

\section{Constraints from flavor changing neutral currents}
\label{Sect:FCNC}

Light DM with a significant Higgs coupling may have implications for low energy phenomenology, in particular, for flavor changing meson decays. 
Indeed, DM can be produced in the course of such decays via the ``Higgs penguin'' diagram. Non-observation of these processes imposes a constraint on DM properties.

Consider first the fermion DM case. The strongest constraints come from flavor changing neutral current (FCNC) processes with missing energy, 
e.g. $ B \rightarrow K \chi \chi$ and  $K \rightarrow \pi \chi \chi$. These are allowed for $m_\chi \lesssim 2.4\;$GeV and $m_\chi \lesssim  180\;$MeV, 
respectively. The relevant flavor changing vertices are the one-loop generated SM effective couplings \cite{Willey:1982mc}
\begin{equation} 
    \bar b s \,h ~,~ \bar s d \,h~,...   
 \end{equation}   
These couplings are due to the $W$-boson exchange with the Higgs boson attached to one of the propagators in the loop. For example,
the $\bar b_R s_L $ coupling has the value (see equation (3.2) of \cite{Haber:1987ua})
\begin{equation}
C_{b_R s_L} \simeq   {3g^3 \over 128 \pi^2 } V_{tb} V_{ts}^* {m_b m_t^2 \over m_W^3} \simeq 6 \times 10^{-6} \;,
\end{equation} 
in terms of the $W$-boson mass $m_W$ and the Cabibbo–Kobayashi–Maskawa matrix elements $V_{ij}$.
The Higgs boson couples to DM via $ {v\over \Lambda} \,h \bar \chi \chi  $ such that, upon integrating it out, we get an effective interaction 
\begin{equation}
\Delta {\cal L}_{\rm eff}^\chi  \sim    {1 \over \Lambda \times 10^7\; {\rm GeV}} \;  \bar b_R s_L \,\bar \chi \chi \, + {\rm h.c.} \,.
\end{equation}
The invisible Higgs decay constraint imposes  $\Lambda \gtrsim {\rm few}\times 10^4\;$GeV such that the above dim-6 operator is suppressed 
by more than $10^{11}\, {\rm GeV}^2$. This is much higher than what is required by non-observation of $ B \rightarrow K \chi \chi$, i.e.~the suppression scale 
squared must be above  $10^{8}\, {\rm GeV}^2$ \cite{Kamenik:2011vy}. Hence, the FCNC bound is trivially satisfied. A similar result applies 
to the $\bar b d \, h$ coupling.

In the $\bar s d$ transition, we have an analogous situation: $C_{s_R d_L}  \sim 2 \times 10^{-9}$ and the suppression scale squared enforced by 
the invisible Higgs decay is at least $\sim  10^{15}\, {\rm GeV}^2$. On the other hand, the corresponding FCNC bound is only $10^{12}\, {\rm GeV}^2$ 
\cite{Kamenik:2011vy}. These considerations also apply to the CP-violating fermion-Higgs coupling $\propto 1/ \Lambda_5$.

Let us now turn to the scalar DM case. Integrating out the Higgs boson, one obtains a dim-5 operator
\begin{equation}
\Delta {\cal L}_{\rm eff}^S  \sim    {\lambda_{hs} \over   10^8\; {\rm GeV}} \;  \bar b_R s_L \,SS \, + {\rm h.c.} \,.
\end{equation} 
The invisible Higgs decay bound requires $\lambda_{hs}\lesssim 10^{-2}$, which makes the suppression scale of the dim-5 operator to be of order $10^{10}\;$GeV.
The FCNC constraint is much weaker, of order $10^7\;$GeV \cite{He:2022ljo}. Analogous considerations apply to the $b-d$ and $s-d$ transitions.
We thus find that meson decays with missing energy do not impose additional constraints on our framework.

\section{Conclusion}
\label{Sect:conclusion}

We have studied the Higgs decay into light DM within the framework of freeze-in DM production at stronger coupling. Both spin 0 and spin 1/2 DM candidates 
with the Higgs portal couplings have been considered and shown to be compatible with the correct DM relic abundance. The required reheating temperature 
$T_R$ does not exceed a few GeV, which is consistent with cosmological constraints. At low temperatures, the hadronic DM production channels, such as 
pion annihilation, become important and are accounted for by relating the corresponding reaction rate to the Higgs decay into hadrons. For MeV-scale 
reheating temperatures, DM is produced predominantly via electron-positron annihilation.

A significant Higgs coupling to DM requires either Boltzmann suppression of DM production or DM to be lighter than the muon. In the former case,
DM remains non-thermal due to $T_R \ll m_{\rm DM}$, while in the latter case thermalization is suppressed by the electron Yukawa coupling.
With both options, the invisible Higgs decay can be very efficient. The current LHC bound already rules out large regions of the parameter space,
while the HL-LHC and FCC can probe the model further, down to percent and sub-percent level of ${\rm BR_{inv}}(h\rightarrow {\rm DM})$.

Altogether, we find good prospects of observing invisible Higgs decay even in simplest DM models. This conclusion crucially depends 
on the reheating temperature, which we leave as a free parameter instead of (traditionally) assuming it to be very high. 
 \\ \ \\
{\bf Acknowledgements.} 
We would like to thank Yuval Grossman and João Paulo Pinheiro for helpful comments.
O.L. is grateful to the Magnus Ehrnrooth foundation for travel support, which has facilitated collaboration on this project. 
A.P.M.~expresses gratitude to the CERN TH Department for supporting scientific visits, which have contributed to the development of the work presented in this article.
V.O. and A.P.M.~are supported by the Center for Research and Development in Mathematics and Applications (CIDMA) through the Portuguese 
Foundation for Science and Technology (FCT - Funda\c{c}\~{a}o para a Ci\^{e}ncia e a Tecnologia), references UIDB/04106/2020 
(\url{https://doi.org/10.54499/UIDB/04106/2020}) and UIDP/04106/2020 (\url{https://doi.org/10.54499/UIDP/04106/2020}). A.P.M. and  V.O. 
are also supported by the projects with references CERN/FIS-PAR/0019/2021 (\url{https://doi.org/10.54499/CERN/FIS-PAR/0019/2021}), 
CERN/FIS-PAR/0021/2021 (\url{https://doi.org/10.54499/CERN/FIS-PAR/0021/2021}) and CERN/FIS-PAR/0025/2021 
(\url{https://doi.org/10.54499/CERN/FIS-PAR/0025/2021}).
V.O. is also directly funded by FCT through the doctoral program grant with the reference PRT/BD/154629/2022 (\url{https://doi.org/10.54499/PRT/BD/154629/2022}).  V.O.~also acknowledges support 
by the COST Action CA21106 (Cosmic WISPers).
A.P.M.~is also supported by national funds (OE), through FCT, I.P., in the scope of the framework contract foreseen in the numbers 4, 5 and 6 of the article 23, 
of the Decree-Law 57/2016, of August 29, changed by Law 57/2017, of July 19 (\url{https://doi.org/10.54499/DL57/2016/CP1482/CT0016}).
R.P.~is supported in part by the Swedish Research Council grant, contract number 2016-05996, as well as by the European Research Council (ERC) 
under the European Union's Horizon 2020 research and innovation programme (grant agreement No 668679). R.P.~also acknowledges support 
by the COST Action CA22130 (COMETA).

\appendix

\section{Sub-MeV dark matter}

\begin{figure}[h!]
    \centering
    \includegraphics[width=0.49 \textwidth]{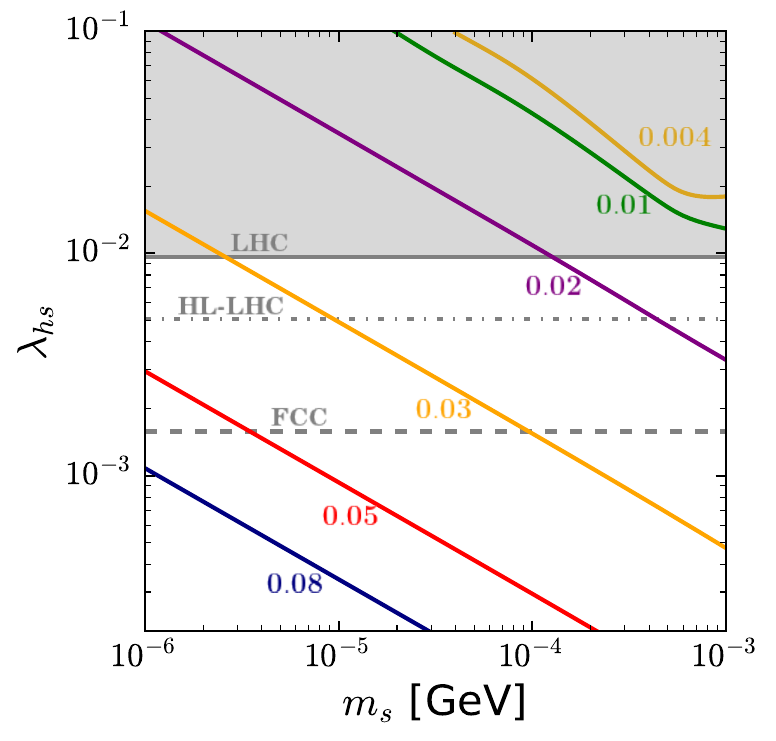}
    \includegraphics[width=0.49 \textwidth]{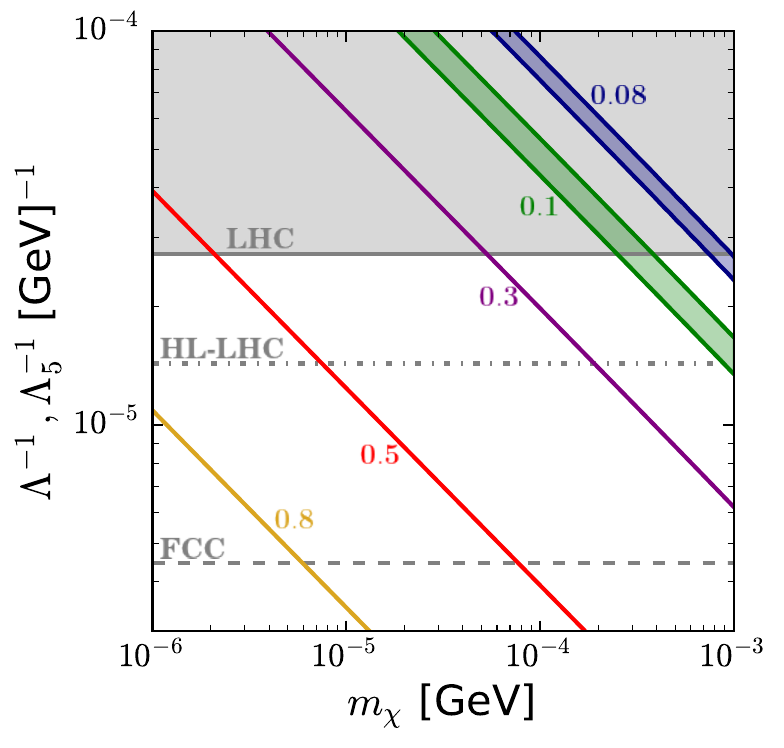}
    \caption{{\it Extrapolation} of Fig.\,\ref{par-space-1} (left) and Figs.\,\ref{par-space-2},\ref{par-space-3} (right) to sub-MeV DM masses with 
    Maxwell-Boltzmann statistics. Along the colored curves marked by $T_R$ in GeV, the correct relic density is reproduced. 
    No structure formation constraint is imposed.
   }
    \label{par-space-kev-s}
\end{figure}

In this work, we have been focused on the DM mass range above an MeV or so. The freeze-in framework, however, allows for lighter DM as well, down to the keV scale.
Our formalism, based on the Maxwell-Boltzmann statistics, cannot be directly applied to this case due to (potentially) significant quantum-statistical effects 
at $T \gg m_{\rm DM}$. A reliable study would require a relativistic approach along the lines of \cite{Lebedev:2019ton,DeRomeri:2020wng}, which makes 
the analysis more computationally intensive. Furthermore, the warm DM constraints have to be taken into account. These depend on the production channel 
and the temperature, and involve the corresponding momentum distribution function.

We reserve a detailed study of the sub-MeV regime for a future work, while in this Appendix we show the results of {\it naive extrapolation} 
of Maxwell-Boltzmann statistics to low masses. Fig.\,\ref{par-space-kev-s} shows the corresponding parameter space down to keV DM masses 
(without the warm DM constraints). These results are tentative and only expected to give the right ballpark for the couplings. In almost the entire 
parameter space, the reheating temperature is far above the DM mass, with low $T_R$ being incompatible with the Higgs data. Nevertheless, 
DM does not thermalize due to the smallness of the Yukawa couplings. In this regime, the initial SM state is much heavier than DM and 
the Higgs-DM coupling increases as DM becomes lighter in analogy with equation \ref{below-mu}. The difference between the CP-even and CP-odd 
cases becomes immaterial in this limit.

\end{document}